\documentclass{jpsj3}
\usepackage{txfonts}
\usepackage{amsmath}%
\usepackage{bm}%
\usepackage{graphicx}
\usepackage{color}
\usepackage{overcite}

\def\gd{GdCu$_2$}
\def\tn{$T_{\rm N}$}
\def\dd{${\delta}$}

\def\qvecf{${\bm q}$={({\dd},~1,~0)}}
\def\qveccf{${\bm q}_{\rm c}$=(2/3,~1,~0)}

\title{Incommensurate Nature of the Antiferromagnetic Order in GdCu$_2$}

\author{Koji Kaneko$^{1,2}$\thanks{koji.kaneko@j-parc.jp}, Chihiro Tabata$^{1,2}$, Masato Hagihala$^1$, Hiroki Yamauchi$^1$, Masato Kubota$^1$, 
Toyotaka Osakabe$^1$, and Yoshichika \=Onuki$^{3,4}$}
\inst{$^1$Materials Sciences Research Center, Japan Atomic Energy Agency, Tokai, Naka, Ibaraki 319-1195, Japan\\
$^2$Advanced Science Research Center, Japan Atomic Energy Agency, Tokai, Naka, Ibaraki 319-1195, Japan\\
$^3$Faculty of Science, University of the Ryukyus, Nishihara, Okinawa 903-0213, Japan\\
$^4$RIKEN Center for Emergent Matter Science, Wako, Saitama 351-0198, Japan\\
} 

\abst{
A magnetic order in orthorhombic GdCu$_2$ was investigated via a single crystal neutron diffraction with thermal neutron.
Magnetic peaks were observed at incommensurate positions described by the ordering vector ${\bm q}$={({${\delta}$},~1,~0)} with {${\delta}$}=0.678 at 4.6~K. 
This ordering vector is close to the commensurate one with ${\bm q}_{\rm c}$=(2/3,~1,~0) reported earlier, but clearly deviates.
The incommensurate nature of the magnetic order in GdCu$_2$ is further corroborated by the peak shift with temperature below $T_{\rm N}$.
}

\begin{document}
\maketitle

Trivalent Gd and divalent Eu ions show unique magnetic properties among rare-earth ions owing to an absence of orbital moment.
Intermetallic compounds with Gd$^{3+}$ and Eu$^{2+}$ can be understood as spin magnetism mediated by RKKY interaction.
Recently, topological magnetic orders, represented by magnetic skyrmion lattice, were discovered in Eu- and Gd-based compounds\cite{Kakihana2018, Kaneko2019, Hirschberger2019a,Kurumaji2019,Khanh2020,Takagi2022}.
Magnetic skyrmions in $f$-electron systems have characteristic properties such as short periodicity and anisotropy, compared with $d$ electron systems, which can result from different formation mechanism.

An orthorhombic compound {\gd} attracts renewed interests in this respect.
{\gd} undergoes an antiferromagnetic transition around 40~K\cite{Luong1985,Koyanagi1998}.
The magnetic structure in the ground state was reported to be a helical one characterized by the ordering vector {\qveccf}\cite{Rotter2000}.
Because of the strong neutron absorption of Gd,  hot neutron, a wavelength of ${\sim}$0.5~{\AA}, was used to minimize attenuation.
On the other hand, a short wavelength made it difficult to achieve high resolution in a momentum $Q$ space.
In order to get detailed insights into the magnetic order in {\gd}, single crystal neutron diffraction experiment was performed using thermal neutron. 

A single crystal sample of {\gd} was grown by the Czochralski method as same as in Ref.~\citeonline{Koyanagi1998}.
The plate-like sample of ${\sim}4{\times}4{\times}1.8~{\textrm{mm}}^3$ with a plane normal to the $b$-axis was used in the present study.
A single crystal neutron diffraction experiment was carried out on the thermal neutron triple-axis spectrometer TAS-2, 
installed at the T2-4 beam port in the guide hall of the research reactor JRR-3 in Tokai. 
Neutrons with a wavelength of 2.359~{\AA} were obtained and analyzed by pyrolytic graphite (PG) monochromator and analyzer.
 A collimation set of open-40'-S-40'-80' was employed with a PG filter before the sample.
 The sample was attached to a cold finger of a closed-cycle refrigerator to have ($h~k$~0) in the horizontal scattering plane. 

\begin{figure}[]
	\centering
	\includegraphics[width=8.5cm]{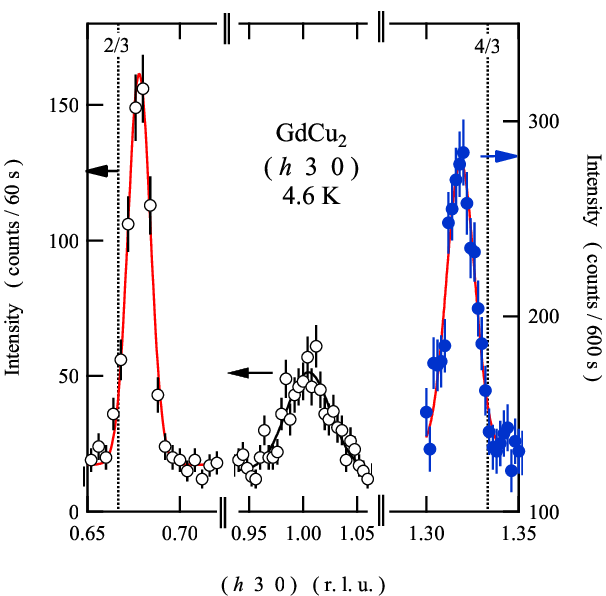}
	\caption{(Color online) Scan along $h$ through (1,~3,~0) collected at 4.6~K below {\tn}. Open circles are plotted with the left axis, while the closed circle uses the axis on the right.
	Dotted lines indicate commensurate peak positions with {\dd}=2/3.}
	\label{prof}
\end{figure}

A linear absorption coefficient for {\gd} with natural Gd is enormously large; roughly 9~${\mu}$m thickness will reduce neutron flux down to 1/$e$.
In order to avoid strong neutron attenuation, reflections close to the reciprocal (0~$k$~0) axis were chosen to have a reflection geometry for the present sample.
Figure~\ref{prof} shows a scan along $h$ through (1~3~0) measured at 4.6~K below {\tn}.
Three peaks were observed, which consists of a nuclear peak at center at (1~3~0) accompanied by satellite peaks on both sides.
Note that the nuclear Bragg peak at (1~3~0) is forbidden for the symmetry $Imma$ .
As the central peak serves as a good reference to determine magnetic satellite peak positions precisely, the PG filter was partly removed and use ${\lambda}$/2 to obtain the peak at (1~3~0).
In smaller $h$ region, a sharp magnetic peak was observed at 0.678(1) whose width corresponds to expected resolution of ${\sim}$0.013~r. l. u..
In contrast, a subtle, but clear magnetic peak was observed in larger $h$ at 1.323(1) plotted in the right axis, which needs 10 times longer counting time to obtain the similar statistics.
By using the single parameter ${\delta}$ to fit the both peak positions with respect to the nuclear peak, ${\delta}$ was determined as 0.678(1).
This is evidently smaller than 2/3, which is indicated by broken lines in the figure.
On the other hand, a scan along $k$ confirms that the magnetic peak is centered at $k$=3, namely the $k$ component is 1.
The results indicate that the magnetic order in {\gd} is not commensurate with {\qveccf}, but has an incommensurate nature with {\qvecf}.

\begin{figure}[ttt]
	\centering
	\includegraphics[width=8cm]{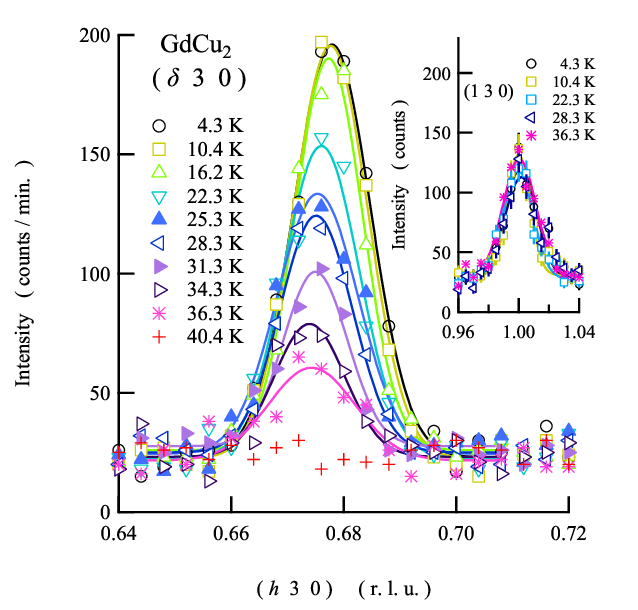}
	\caption{(Color online)Representative line scan profiles along $h$ through ({\dd}~3~0) measured at several temperatures below {\tn}. The inset shows line scan profile through nuclear peak at (1~3~0) recorded at selected temperatures. }
	\label{tprof}
\end{figure}

This finding is also supported by its temperature dependence. 
Figure~\ref{tprof} shows the scan along $h$ through (${\delta}$~3~0) collected at several temperatures below {\tn}.
With increasing temperature, the magnetic peak at the incommensurate position at $h$=0.678 becomes weaker, and vanished at 40.4~K just above {\tn}.
Concomitantly, the magnetic peak exhibits gradual peak shift to smaller $h$ as temperature increases.
This is contrast to the adjacent nuclear peak 1~3~0, which stays at the same position as shown in the inset of Fig.~\ref{tprof}.

\begin{figure}[!ttttt]
	\centering
	\includegraphics[width=8cm]{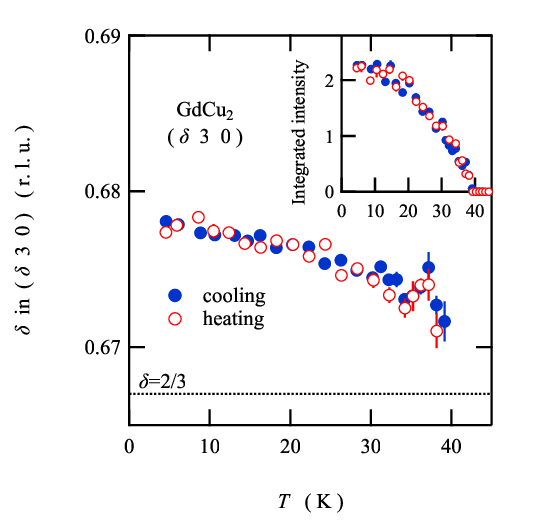}
	\caption{(Color online)Magnetic peak position {\dd} in ({\dd}~3~0) as a function of temperature measured upon both heating and cooling. The inset shows temperature variation of the integrated intensity of the same magnetic reflection.}
	\label{tparam}
\end{figure}%
Figure~\ref{tparam} summarizes the temperature variation of the peak position and intensity obtained by a Gaussian fitting.
Concerning the peak position, the peak at 0.678 at 4.6 K exhibits monotonous decrease with the temperature, and becomes 0.671 at 38.2~K.
The peak positions {\dd} in the entire temperature range are larger than 2/3, as indicated by the broken line in the figure.
No visible difference exists between upon cooling and heating.
A change of the peak position with temperature also supports an incommensurate nature of the transition in {\gd}.
The magnetic peak intensity shown in the inset develops gradually with decreasing temperature.
No visible hysteresis was observed as in the peak position, consistent with a second-order nature of the magnetic transition in {\gd}.

The present results reveal the incommensurate nature of the magnetic order in {\gd}.
The deviation from the commensurate order is small, 0.01~r.l.u. along $a$.
Within the previous magnetic structure model, a small change in periodicity results in a  slight reduction in pitch angle; the angle between neighboring magnetic moments along $a$ is reduced from 120$^{\circ}$ to ${\sim}116^{\circ}$. 
The lock-in transition to commensurate structure could not been seen in the present temperature range down to 4.6~K.

In general, neutrons with a short wavelength, 1~{\AA} or less, are used to study high neutron absorbing material as absorption cross section follows so-called 1/$v$ low, where $v$ corresponds to speed of neutron.
Indeed, the high neutron absorption cross section of Gd, 58500 barns at a typical thermal-neutron wavelength 2.36~{\AA}, can be suppressed to roughly 1/100 at hot neutron region at 0.5~{\AA}.
This suppression of neutron absorption enables quantitative structure analysis as in the previous report \cite{Rotter2000}.
On the other hand, a short wavelength tends to result in a poor resolution in the $Q$ space, in particular for low $Q$ region.
In other words, this setup is not ideal to determine precise periodicity as mentioned in  Ref.~\citeonline{Rotter2000}.  
Whereas an accurate absorption correction and structural analysis could be problematic,
a peak position and its variation with an external parameter can be tracked accurately with thermal neutron.
A complementary use of both short- and long-wavelength neutrons leads to obtain a detailed picture of magnetic order.

In summary, the present study revealed that the antiferromagnetic order in orthorhombic {\gd} can be described with the incommensurate ordering vector with {\qvecf} with {\dd}=0.678 at 4.6~K.
The gradual change of {\dd} with the temperature also supports an incommensurate nature of the transition in {\gd}.

\begin{acknowledgment}
We thank Y. Shimojo and M. Sasaki for their support on neutron scattering experiments. 
This work was supported by JSPS KAKENHI Grants Nos. JP20H01864, JP21H01027, JP21H04987, and No. JP19H04408. 

\end{acknowledgment}

\bibliographystyle{./jpsj}
\bibliography{GdCu2}

\end{document}